# The Blind Men and the Elephant: Mapping Interdisciplinarity in Research on Decentralized Autonomous Organizations

Giorgia Sampò
Digital Democracy Center
University of Southern Denmark
gisa@sam.sdu.dk

Oliver Baumann
Department of Business and Management
University of Southern Denmark, Denmark
oliv@sam.sdu.dk

Marco Peressotti
Department of Mathematics and Computer Science
University of Southern Denmark
peressotti@imada.sdu.dk

**Abstract:** Decentralized Autonomous Organizations (DAOs) are attracting increasing interest from various disciplines, particularly business, economics, and computer science. However, much like the parable of the blind men and the elephant, where each observer sees only part of the phenomenon, DAO research has largely remained fragmented across disciplines, limiting a comprehensive understanding of their potential. This paper investigates the extent DAO scholarship has achieved meaningful interdisciplinary integration. We address this question through an analysis of knowledge flows between Business & Economics and Computer Science, using citation network analysis, topic modelling, and outlet analysis. We find that while DAOs generate vibrant interdisciplinary discourse, the interactions remain predominantly applied and case-driven, with limited theoretical integration. By mapping interdisciplinary exchanges, we highlight key gaps and opportunities for greater synthesis across fields. Strengthening the alignment between organizational and technical insights is crucial for advancing DAO research and fostering a more cohesive interdisciplinary framework.

**Keywords:** Decentralized Autonomous Organizations; Blockchain; Interdisciplinarity; Business and Economics; Computer Science.

## 1. Introduction

Decentralized Autonomous Organizations (DAOs) are attracting attention from scholars across disciplines, particularly in organization science and computer science. While DAOs share the core features of other organizations (Ellinger et al., 2024; Hsieh, 2018; Puranam et al., 2014), they fundamentally differ in their governance models due to their lack of human executives to address challenges of coordination and cooperation (Foss and Xu, 2023). Instead, they rely on blockchain-based solutions and a “code is law” logic whereby agreements – the so-called “smart contracts” – are written and executed as computer code (Buterin et al., 2014).

In doing so, DAOs simultaneously draw upon and challenge existing theories of organizational governance while also pushing the boundaries of blockchain technologies and algorithmic coordination. As a novel form of decentralized governance, DAOs blur the line between human and algorithmic decision-making, making them a natural subject of interdisciplinary research: organizational innovations in DAOs are shaped by (blockchain) technology, whose specific uses and developments are driven by the distinct objectives that these organizations pursue (Amend et al., 2024; Ellinger et al., 2024; Lumineau, et al., 2021; Morrison et al., 2020).

However, despite this mutual relevance of organization science and computer science, our understanding of the interdisciplinary nature of research on DAOs remains limited. Take, for instance, smart contracts. Computer scientists emphasize their technical challenges and potential capabilities (Buterin, 2014; Khan et al., 2021; Zheng et al., 2024), whereas organizational scholars focus on the innovative governance mechanisms that smart contracts introduce (Ellinger et al., 2024; Gregory et al., 2024; Murray et al., 2019). At the same time, integration between these two fields appears to be limited: while business scholars tend to possess only a superficial understanding of the technical aspects of smart contracts, computer scientists often overlook the applications envisioned by their organizational counterparts.

This gap is noteworthy because DAOs reflect how organizations and technologies might co-evolve. Without a comprehensive understanding of the challenges and shortcomings of existing solutions, knowledge remains siloed, potentially precluding researchers from envisioning more effective organizational designs. Existing approaches resemble the parable of the blind men and the elephant: in the story, a group of blind men encounter an elephant for the first time, each of them only touching a part of the animal before describing it to the group. This fragmented perception leads them to mistake their partial view for the entire phenomenon, creating a distorted understanding of the whole. Similarly, differences in disciplinary perspectives often lead to misalignment, hindering the development of coherent and comprehensive models.

Although the need to integrate insights from various fields is widely recognized, DAO scholars, much like the blind men, struggle to develop a comprehensive interdisciplinary framework (Bonnet and Teuteberg, 2024). In some versions of the story, the blind men begin doubting each other’s reliability, leading to conflict, while in others, they collaborate, piecing together their perceptions to “see” the whole elephant. Yet, another version introduces an external observer, who, having broader vision of the animal, enlightens them on its appearance. Just as this observer unifies perspectives into a single understanding of the elephant, informing scholars about the state of interdisciplinary research is crucial to guiding future work towards integrative approaches, meaningful collaborations, and theoretical advancements. Given the growing prominence of DAOs in both academic research and practical applications, it is critical to assess whether DAO research has reached a level of genuine integration or remains largely fragmented.

True interdisciplinary approaches require scholars to collaboratively build causal structures and unified theoretical and critical perspectives (Frodeman, 2015; Klein, 2015; Klein 2017), juxtaposing shared knowledge to provide a complete description of the phenomenon under analysis. Unlike the blind men in the story, scholars are generally aware of their limited perspective but may still struggle to construct a unified understanding. Like the observer in the story, our work aims to guide this effort – not by providing a complete description of DAOs, but by mapping existing interdisciplinary exchanges to facilitate and direct future collaboration. By systematically mapping how knowledge flows between disciplines, we aim to identify whether a shared conceptual foundation for DAO research is emerging – or if disciplinary silos continue to dominate. We aim to contribute not only to the literature on DAOs but also to broader discussions on how emerging technologies drive shifts in governance and organizational design.

To assess the maturity of interdisciplinary research on DAOs, we analyze knowledge flows between Business and Economics and Computer Science. Our analysis follows three steps and leverages conceptual frameworks and quantitative methods for measuring interdisciplinarity (Glänzel and Debackere, 2022), using citations network analysis, outlet analysis, and topic modeling. First, we conduct a citation network analysis to identify "boundary papers" – works present in both fields – and examine patterns of interdisciplinary interaction (Karunan et al, 2017). This includes assessing whether cross-field exchanges occur via direct citations or through a shared knowledge space, as well as analyzing the diffusion of concepts and mutuality of interactions. Next, we apply topic modeling to explore the key themes driving interdisciplinary discussions, investigating whether organizational and technical research share common conceptual ground or remain discipline specific. Finally, we employ network analysis to investigate relationships between authors and publication outlets, which cannot be captured by citation networks. This multi-method approach allows us to uncover not just *where* but also *how* interdisciplinary exchanges occur.

Our analysis confirms the inherently interdisciplinary yet fragmented nature of DAO research. We find that while DAO research has generated vibrant interdisciplinary discussions, these interactions are heavily applied, with limited theoretical contributions. Moreover, the community driving this interdisciplinary research remains fragmented, with few journals and conferences serving as aggregation hubs, and connected by a small number of scholars. As a result, DAOs lack robust models and frameworks capable of fully capturing their interdisciplinary potential and supporting integrated research approaches. Strengthening the alignment between organizational and technical insights will not only advance theoretical understanding but also help shape best practices for designing and governing the next generation of DAOs. Moreover, interdisciplinary research has also the potential to bridge and even transform the original disciplines it draws from (Pacheco et al., Klein, 2017), fundamentally reshaping domain-level knowledge creation.

Our paper is organized as follows: Section 2 provides a conceptual background on DAOs and interdisciplinary research and discusses relevant related work. Section 3 presents our methodology, outlining each step of the analysis and its contribution to understanding interdisciplinary interactions in DAOs. Section 4 presents our results, while Section 5 concludes with a discussion of our results and directions for future research.

## 2. Conceptual background

### 2.1. Decentralized Autonomous Organizations[1]

[1] A set of definitions of common terms in the blockchain domain can be found in Appendix A.

DAOs are consensus-based blockchain communities designed to coordinate economic activity, enabling decentralized and reliably coordinated decision-making using blockchain technology and smart contracts. As such, they simultaneously constitute the building blocks for the new forms of economic governance enabled by blockchain based technologies and the agents concretizing it (Davidson et al., 2018). DAOs offer several governance benefits, including "ease of asset allocation, streamlined group decision-making, increased empowerment and inclusion of regular members, and a transparent and trustless nature" (Jirasek, 2023). These features characterize DAOs as novel organizational forms that seek to address the universal problems of organizing (Puranam et al., 2014) by leveraging blockchain technology. Their unique traits, such as core developers assuming traditional management roles and relying exclusively on community-driven democratic processes for decision-making, transaction verification, and public ledger updates (Hsieh, 2018), make them particularly interesting for organizational scholars.

Acting as an architectural innovation (Lumineau et al, 2021), blockchain recombines existing concepts in previously unseen ways, thereby enabling DAOs to facilitate consensus and organize collaboration among members. By promoting full transparency and decentralized, rule-based governance structures, DAOs can potentially reduce agency costs, align stakeholder interests, and foster distributed consensus (Li et al., 2024). This is achieved through a "code is law" approach, wherein smart contracts are encoded on the blockchain and automatically executed when predefined conditions are met, granting participants adherence to rules, agreements and governance (Singh and Kim, 2019). Such automation minimizes reliance on formal or informal contracts, distinguishing DAOs from traditional forms of decentralized organizations (Amend et al., 2024).

Furthermore, DAO decision-making processes are based on the use of tokens to manage governance rights, often requiring intricate balancing to ensure stakeholder representation while preventing centralization of power (Li et al., 2024). Governance within DAOs uniquely merges corporate governance with IT governance, making the two inseparable (Morrison et al., 2020), a distinction enabled by blockchain technology. Nonetheless, DAOs can adopt various voting mechanisms, ranging from a "one-person-one-vote" model to a "one-token-one-vote" approach, akin to stakeholders' rights in companies (Feichtinger et al, 2023; Sims, 2019), with numerous hybrid structures falling in between.

The decentralization and automation offered by DAOs present both opportunities and risks. On one hand, they could empower communities to create egalitarian ecosystems based on a one-person-one-vote model, eliminating central authority and control and single points of failure (Chotkan et al. 2022). On the other hand, they can serve as tools to obscure and legitimize the decisions of a ruling elite under the guise of community governance (Davidson, 2024; Feichtinger et al., 2023; Liu, 2024). Their characteristics position DAOs as one of the most innovative applications of blockchain, with the potential to develop a robust business model encompassing key components such as partnerships, activities, resources, cost structures, revenue streams, channels, customer segments, and ultimately, value propositions (Saurabh et al., 2023). However, DAOs function within a dynamic ecosystem in which traditional governance paradigms are continuously evolving and being reinterpreted. As a result, they remain subject to the practical limitations of agency theory, which requires the implementation of checks and balances to prevent excessive concentration of power and ensure the alignment of diverse stakeholder interests (Alawadi et al., 2024).

As a relatively new phenomenon, DAOs present numerous unanswered questions regarding their challenges and applications, both from a technical and an organizational perspective, leading to an underutilization of their full potential (Yu et al, 2023). Researchers still need to tackle the specific challenges of achieving effective decentralization, given the complexities of protocol governance, to avoid replicating traditional corporate models without critical adaptation (Foss and Xu, 2023), to tackle the issue of passive members and to mitigate power imbalances (Peña-Calvin et al., 2024).

The distinct nature of this emerging organizational model necessitates the development of specific frameworks rooted in an interdisciplinary perspective: understanding the organizational novelty of DAOs requires an integrated approach that considers both their computational dimensions and the theoretical insights provided by organizational scholarship.

## 2.2. Interdisciplinarity

Innovation increasingly occurs at the boundaries of disciplines, moving beyond the traditional "discipline within a department" structure towards interdisciplinary approaches (Pfirman and Martin, 2017). Fields such as molecular biology, social psychology, or digital humanities have evolved from loosely connected concepts to fully integrated paradigms. Some have even developed rich frameworks reflecting on the needs, methods, and benefits of interdisciplinarity (McCarty, 2013; Klein, 2015; Klein, 2017).

This evolution distinguishes interdisciplinarity from multidisciplinary. While multidisciplinarity involves juxtaposing inputs from separate disciplines, interdisciplinarity integrates information, methods, tools, and theories from two or more fields to create a unified understanding (Klein, 2015, Klein 2017). Multidisciplinarity promotes breadth and diversity in knowledge creation, whereas interdisciplinarity focuses on blending and linking disciplinary inputs to address complex questions or problems holistically, whether by individuals or teams (Klein, 2015; Klein 2017; Pacheko et al., 2017).

Interdisciplinary interactions not only generate new ideas and approaches but also reshape how traditional disciplines frame knowledge (Klein, 2017; Pacheko, 2017; Pfirman and Martin, 2017). They create "trading zones" and "creoles" for scholars to exchange, debate, and refine interdisciplinary concepts (Galison, 1997; Klein, 2015). These spaces enable the expertise-sharing necessary for interdisciplinary fields to progress (Galison, 1997; Klein, 2015; Klein, 2017) and are fundamental for the emergence of a singular domain from the admixture of different knowledge streams.

The level and purpose of interdisciplinary interactions influence the structure and impact of knowledge exchanges. For instance, *methodological interdisciplinarity* improves outcomes by borrowing methods and concepts, while *theoretical interdisciplinarity* aims at integrating knowledge at the theoretical level, thus creating new frameworks and theories. *Instrumental interdisciplinarity* aims at innovating a product or meeting a practical need, whereas *critical interdisciplinarity* challenges and transforms dominant knowledge structures (Klein, 2015; Klein, 2017). While interdisciplinary research often emphasizes case studies and practical approaches (Frodeman, 2015), only theoretical and critical interdisciplinarity can establish new paradigms, bridging gaps between domains and fostering groundbreaking insights.

Scholars have studied knowledge exchanges through both qualitative and quantitative approaches, aiming to classify and understand interdisciplinary interactions (Frodeman, 2015; Klein, 2017). Bibliometric methods and network metrics are particularly well-suited for analyzing scholarly interactions quantitatively, as they can account for both cognitive (knowledge flows) and organizational (real-life collaborations) dimensions of interdisciplinarity (Glänzel and Debackere, 2022; Karunan et al., 2017). In studying DAOs, these methods can help capture how blockchain-oriented research and organizational theories merge or diverge, which is crucial for building a more unified understanding of decentralized governance.

However, bibliometric methods often struggle to capture the social dynamics that drive knowledge integration (Wagner et al. 2011). Citation network analysis addresses these limitations by exploring interdisciplinary interactions, identifying shared knowledge and characterizing connections between domains (Karunan et al., 2017). By pinpointing which papers and authors most frequently cross boundaries, we can see where true integration begins to form, while by highlighting the impact of cross-disciplinary connections and boundary knowledge through a tailored framework and analyzing the role of authors and venues of publication, we will be able to assess the structure and nature of interdisciplinary interactions. The additional focus on the themes of such interactions, provides us with a deeper understanding of connections between fields and should allow us to overcome the limitations of traditional bibliometric approaches.

### 2.3. Related work

Recent research on DAOs has primarily focused on governance models, with growing interest in addressing power imbalances, token distributions, and (de)centralization dynamics. While these studies share a common focus, their approaches vary across disciplines. Some research is theory-driven, assimilating DAOs into existing models and frameworks (Saurabh et al., 2023) or developing new ones within specific disciplines, (Davidson, 2024), while others adopt inductive approaches, drawing on practitioners' insights (Alawadi et al., 2024). A third category applies quantitative methods to analyze voting power, decentralization levels, and community activity (Feichtinger et al., 2023; Peña-Calvin et al., 2024). Similarly, another branch of literature focuses on developing new governance models to better exploit DAOs' potential. Some scholars explore how DAOs integrate with the Web3 ecosystem (Chotkan et al., 2022; Yu et al., 2023), while others compare decentralized governance to existing frameworks (Li and Chen, 2024). Further studies introduce cooperation and reputation metrics designed specifically for tokenomics (Nasrulin et al., 2022), offering alternative mechanisms for DAO governance.

Comprehensive analyses have also been conducted through literature reviews on blockchain governance. By examining existing governance models at the blockchain level, where DAOs are the most prominent instantiation of distributed ledger-based communities, scholars have identified key shortcomings and possibilities in the field. Research has highlighted DAOs' potential in combating corruption and enhancing transparency (Ibrahimy et al., 2023) and has helped identifying the foundational elements that distinguish blockchain governance from traditional IT governance (Morrison et al., 2020), addressing its design, adoption, and enforcement models (Liu et al., 2022).

Additionally, literature reviews focused specifically on DAOs have sought to provide tailored definitions and analyze the challenges and opportunities offered by this emerging technology. Some scholars have examined the benefits and costs of leveraging Web3 architecture, addressing both organizational and technical requirements for decentralization and self-governance (Yu et al., 2023). Others have critically analyzed the meanings and implications of “decentralization” and “automation,” taking a comprehensive approach across multiple disciplines and emphasizing the need for hybrid governance models (Bonnet and Teuteberg, 2024).

Beyond theoretical discussions, scholars have developed taxonomies to organize existing knowledge about DAOs. These classifications are primarily based on scholarly publications (Bonnett and Teuteberg, 2024), inductive insights from DAO implementations (Ziegler and Welpe, 2022; Wang et al., 2019), and practitioners' domain expertise (Alawadi et al., 2024; Axelsen et al., 2022).

Despite these efforts, research has yet to adopt a holistic perspective. Existing studies recognize DAOs' interdisciplinary nature, spanning legal, computational, organizational, and social domains, but fail to provide a unifying framework that bridges these domains. While some aspects of DAOs can be examined in isolation, their technological and organizational dimensions are tightly interwoven – and cannot be fully understood separately.

While existing reviews encompass knowledge from multiple disciplines – including computer science and business and economics – they continue to treat technical and organizational factors separately. Moreover, existing taxonomies lack connections between the technical aspects of DAOs and the governance aspects they relate to. To the best of our knowledge, no published research on DAOs explicitly examines the interdisciplinary nature of DAOs or systematically integrates knowledge from multiple sub-domains. This gap highlights the need for a comprehensive framework that bridges the organizational and technical perspectives necessary for a deeper understanding of DAOs.

## 3. Methods

### 3.1. General approach

This study integrates knowledge from the fields of Business and Economics (BE) and Computer Science (CS), to identify and analyze knowledge flows between scholarly communities. Our goal is to examine the role and forms of interdisciplinarity that characterize the conversation on Decentralized Autonomous Organizations and to establish a foundation for consistent engagement between scholars from different disciplines. A casual review of existing literature reveals that multiple definitions of DAOs have been proposed within each academic community, no unified definition has yet emerged. Given the nature of DAOs and the innovation they introduce in both fields, we argue that analyzing the relationship among papers at the intersection of BE and CS – specifically, interdisciplinary interactions and the ways in which they shape and are shaped by existing knowledge – is essential for establishing a robust conceptual framework, which can provide a foundation for further exploration of the topic.

Our study takes a quantitative approach[2], using bibliometric analysis of citation networks to assess the existence and structure of connections among scholars from different disciplines. By analyzing the papers emerging as relevant in these networks, we seek to understand whether and how the two communities interact, and the processes through which new knowledge is generated. Furthermore, by applying topic modeling and examining publication outlets, we identify key journals, conferences, and scholarly conversations where DAO-related knowledge is being developed and refined.

### 3.2. Data collection and data cleaning

Our bibliometric analysis is based on data from Scopus and Web of Science (WoS). First, we queried both databases through their APIs, using the keywords listed in Appendix B[3]. In Scopus, searches targeted the “Title”, “Abstract”, and “Keywords” fields, while in WoS, we used the “Topic” field to ensure the closest match across platforms. To maintain consistency in disciplinary scope, we limited searches to “Social Sciences” and “Computer Science” in Scopus, while in WoS, we included papers from “Business”, “Economics”, “Management”, and “Multidisciplinary Sciences”, and several subfields of “Computer Science” (Hardware & Architecture, Information Systems, Interdisciplinary Applications, and Theory & Methods).

These restrictions in research fields and keywords were implemented to exclude irrelevant results, such as studies on D-amino acid oxidase (DAO) in biology. However, we intentionally maintained a

[2] The python code employed for the analyses in this article can be found at https://doi.org/10.5281/zenodo.14867817

3 Queries were last ran in July 2024

broad perspective and did not limit our scope strictly to “Business” and “Computer Science”. Finally, the collected papers were divided into two main datasets, reflecting the different research perspectives in our integrative analysis.

Data cleaning was conducted in three steps: (i) removing papers that were clearly unrelated to the topic, such as those using “DAO” as an alternative spelling of “TAO” or referring to Data Access Objects; (ii) selecting only papers that focus on DAOs instantiations, theoretical or technical foundations, precursors (e.g., decentralized online communities in Web 3.0) or blockchain-specific applications; and (iii) removing papers missing essential metadata such as DOIs.

We started by querying Scopus API, employing the keywords and queries reported in Annex. 1 - Keywords. The resulting datasets, one for BE and one for CS, contained information about the retrieved papers, including author(s), title, date of publication and DOI. This first search provided us with a total of 3369 papers (single research yielded 1286 papers for BE and 2707 for CS), that we then manually reviewed and cleaned. After a first screening our dataset was reduced to a total of 1873 papers; during the second round of cleaning, we further reduced our dataset to 571 papers.

Once the cleaning of papers gathered from Scopus was completed, we performed the same research in WoS. This search yielded 790 documents: of these, only 124 were not already included in our results. We therefore proceeded to clean the new subset through the removal of papers that qualified either irrelevant or missing fundamental data (i.e., DOIs), as described above. The final set was composed of 23 papers, that were then added to the bigger set of material gathered from Scopus.

At the completion of all cleaning procedures, our dataset therefore consisted of 594 papers in total: of these papers, 320 belonged to the Business domain, 426 to Computer Science and 152 were present in both fields. It must be noted that papers from WoS, due to the platform labeling logic, belonged to either one category or the other and did not influence in any way the number of common papers.

### 3.3. Citation networks building

We constructed three directed citation networks: one for each discipline (BE and CS) and a combined network incorporating papers from both fields. In all three networks, each node represents a paper, and each edge represents a citation from one paper to another within the network. The initial set of nodes included those retrieved through our search, along with those cited by that initial set.

Citation networks for BE and CS were built using the Scopus Abstract API by retrieving citation data for every paper in our dataset. We then refined the networks by removing isolated nodes (i.e., those with on incoming or outgoing citations) and nodes with degree of one (i.e., those with only a single citation in either direction). Publications with only one citation were likely included in the initial dataset due to an apparent but weak correlation with the topic but are not very relevant in the final network nor informative about internal connections. These nodes often appear in the network as “leaves”, connecting to the graph through a single connecting paper that cites multiple such nodes. This pattern suggests that while the connecting paper relates to DAOs, the cited papers belong to a different scholarly community (e.g., e-voting in political science). Removing these leaves eliminates noise while preserving papers with multiple connections to the network, preventing the overrepresentation of unrelated topics (e.g., e-voting or smart-city governance). Also, only papers reporting a DOI were employed for the construction of the networks, thus excluding “gray literature” from our analysis while allowing us to uniquely identify each of the nodes in our network, without any possible duplication or homonymy.

The final networks consisted of 1,311 nodes and 3,617 edges for CS and 1,084 nodes and 3,017 edges for BE. We then merged the two networks to form a combined citation network, which served as the basis for further analysis. From this merged network, we then identified "boundary papers" – those at the intersection of the BE and CS datasets – resulting in a total of 501 nodes and 1193 edges.

Descriptive statistics regarding the final networks are reported in Table 1: outdegree corresponds to the number of citations made, while indegree corresponds to citations received, on average, by each paper. The mean of citations made by each paper (their references) is lower than one would expect, but quite consistent between networks. This is both due to the removal of leaf nodes and the decision to exclude works that do not have a DOI[4]. Moreover, given our data collection strategy, excluding works missing a DOI, the value of our citations does not represent the overall number of citations gathered by the referenced paper, but the number of citations made from the other works involved in our study. The table therefore provides an indication of the relative weight of each paper in our analysis and cannot be interpreted as an overall indication of the importance of the works reported.

**Table 1: Network statistics**

| Network | Measure | Mean | Median | Mode | Min | Max | Standard deviation |
|---|---|---|---|---|---|---|---|
| Business & economics | Outdegree | 10.7 | 7 | 4 | 1 | 71 | 10.7 |
| | Indegree | 3.4 | 2 | 2 | 1 | 50 | 3.8 |
| Computer science | Outdegree | 9.9 | 7 | 4 | 1 | 59 | 9.0 |
| | Indegree | 3.4 | 2 | 2 | 1 | 44 | 3.5 |
| Boundary papers | Outdegree | 7.1 | 5 | 4 | 1 | 36 | 6.0 |
| | Indegree | 3.1 | 2 | 2 | 1 | 26 | 3.1 |
| Merged network | Outdegree | 11.2 | 8 | 4 | 1 | 71 | 10.3 |
| | Indegree | 3.7 | 2 | 2 | 1 | 69 | 4.8 |

### 3.4. Cross-citations and boundary papers analysis

Next, we analyzed interdisciplinary connections using the framework proposed by Karunan et al. (2017). First, we measured the strength of boundary papers (denoted as *IDp*) between the two citation networks (BE and CS) using the following formula:

$$IDp = \frac{n|k_1}{nc|k_1}$$

where:

- $n|k_1$ is the number of shared nodes (papers) between the two networks that have at least one citation (either incoming or outgoing), filtering out uncited papers.
- $nc|k_1$ is the total number of nodes in the merged network, considering only those with at least one citation.

4 Publications without a DOI can include, but are not limited to, blogposts, newspaper articles, book chapters, and other forms of grey literature

This metric provides an accurate representation of scholarly interaction and influence through papers that intersect both disciplines.

To assess the level of direct interaction between the two fields, specifically through cross-disciplinary citations, we used the following formula:

$$IDl = 1 - \frac{m1 + m2 - mb}{mc}$$

Where:

- $m1$ and $m2$ represent the number of citations in each of the two separate networks
- $mb$ is the number of common citations between both networks.
- $mc$ is the number of citations in the merged network.

To determine the number of cross-disciplinary references, we apply:

$$m = mc - (m1 + m2 - mb)$$

where *m* represents the number of citation edges linking papers across the two disciplines.

These formulas quantify the strength of interdisciplinary interaction by through direct citations. Specifically, *IDl* measures the proportion of interdisciplinary citations, adjusting for common edges in both networks. A higher *IDl* value indicates stronger interdisciplinary connections. Subtracting $mb$ in the numerator ensures that shared citations, which do not strictly indicate cross-disciplinary, do not inflate the measure of interdisciplinary interaction.

The final step in analyzing the strength of interdisciplinary interactions involves calculating the Dominance Level (*D*) using:

$$D = \frac{IDp}{IDl}$$

Where:

- $D$ represents the Dominance Level.
- $IDp$ measures interdisciplinary interaction based on common boundary papers.
- $IDl$ quantifies interaction strength through cross-disciplinary citations.

The interpretation of $D$ is as follows:

- If $D > 1$, then $IDp > IDl$, indicating that interdisciplinary interaction is primarily driven by shared nodes (boundary papers), suggesting stronger internal interlinkage within the disciplines.
- If $D < 1$, then $IDl > IDp$ , meaning interaction is dominated by cross-disciplinary citations, highlighting stronger external exchanges between the disciplines.

Finally, as an additional measure of interdisciplinary engagement, we calculated the *linkage factor* [5]Karunan et al., 2017), a measure to assess the presence, direction, and balance of mutual contributions between the two disciplines.

[5] The *linkage factor* from network 1 to network 2 is the ratio of the net outflow of knowledge from network 1 to network 2 to the net outflow of knowledge from network 2 to network 1. As outflow from a vertex can be expressed in

### 3.5. Mode of interaction analysis

The framework then examines the *kinds of emergence* (Karunan et al., 2017) observed among boundary papers, which help characterize knowledge flows between fields and providing insight into the most relevant forms of interdisciplinary interaction. Identifying the dominant modes of interaction allows for a qualitative understanding of how interdisciplinary interactions shape overall connections between fields. While all types of emergence involve boundary papers and share a common ground between disciplines, they differ significantly in how they shape interdisciplinary knowledge flows. Below, we outline the different kinds of emergence and their interpretation:

1. *Interdisciplinary diffusion*: The boundary paper serves as a bridge for knowledge transfer between fields (i.e., a paper from field A cites a boundary paper, which in turn cites a paper from field B, or vice versa; A→*boundary*→B). These papers act as "*diffusion channels*", facilitating the translation and integration of knowledge between disciplines, making it more accessible and applicable to both fields.
2. *Circular diffusion*: The boundary paper links nodes within the same field (i.e., a paper from field A cites the boundary paper, which also cites another paper from field A; A→*boundary*→A). These papers function as "*field enhancers*", channeling knowledge back into their original field after adding interdisciplinary insights.
3. *Cross fertilization emergence*: The boundary paper synthesizes knowledge from both fields (i.e., it cites papers from both disciplines; A←*boundary*→B). These papers act as "*field convergents*", blending and integrating domain-specific knowledge from both areas to create a novel interdisciplinary perspective.
4. *Boundary triggered emergence*: The boundary paper is cited by papers from both fields (i.e., it serves as a reference point for researchers in both domains; A→*boundary*←B). These papers act as "*divergence triggers*", generating interdisciplinary themes that may lead to the development of new fields. Their perspectives are relevant across disciplines but extend beyond the primary scope of either field.
5. *Interdisciplinary emergence*: The boundary paper cites another boundary paper (*boundary→boundary*). These papers act as "*boundary extension papers*", fostering interdisciplinary themes that are potentially further extended by subsequent research in both sub-fields, but substantially developing on their own.

### 3.6. Topic extraction

We constructed five different networks by grouping papers based on their interaction modes. To identify communities within these networks, we applied the Louvain modularity algorithm (Blondel et al., 2008), a greedy optimization method designed to detect non-overlapping communities in large networks. For each identified community, we then aggregated the titles and abstracts of its papers, and extracted the main topic covered. To mimic *close reading*, or more specifically the human activity of extracting, aggregating and defining concepts from a pool of texts associated with close reading, and to ensure human-readable results, we used GPT O1[6] for topic extraction instead of traditional topic modeling methods. Topic models represent statistical distributions of co-occurring words in a

---

terms of the indegree of vertex, sum of indegrees of all the vertices in network 1 expresses the net outflow of knowledge from the set of vertices that represent network 1. Similarly, sum of indegrees of all vertices in network 2 expresses the net outflow of knowledge from network 2 (Karunan et al., 2017).

[6] The prompt used for GPT-based topic classification is provided in Appendix C.

set of documents and consequently need further interpretation to assign topics a consistent description and naming: on the other hand, an LLM such as GPT 0.1 can produce the kind of explanation required in our analysis, in the form of human readable labels attached to each community of papers.

To ensure reliability of results, we ran consistency tests, by having the AI analyze the same network different times, and by changing the order in which the networks were presented to the API. While some labels changed their formulation, either slightly (e.g., token economics/tokenomics, DAOs/decentralized autonomous organization) or substantially (e.g., Blockchain security and risks/blockchain application and security, Blockchain in crowdsourcing and social systems/Blockchain for governance and resource management), many remained mostly unchanged. Moreover, and most importantly, the qualitative nature of the labels was not affected, nor was the result deemed inferior to inter-annotators agreement for similar tasks[7].

Next, we collected data on the venues and conferences where boundary papers were published using the Scopus API. To ensure consistency, we aggregated entries referring to the same outlet (e.g., the same conference in different years or journal titles with and without acronyms). We then identified the most frequent publication outlets for boundary papers. Proceedings from conferences belonging to the same associations (e.g., IEEE and ACM), were further grouped under the label "[group name] aggregated proceedings", when individual conferences had only one or two entries in the dataset. Furthermore, by integrating data from previous steps, we analyzed the most represented topics in the largest venues.

### 3.7. Authors and outlet analysis

Finally, we constructed two additional networks: one for venues and one for authors. In the venues network, nodes represent publication venues, with edges connecting venues where the same author has published. In the authors' network, nodes represent authors, with edges linking those who have published in the same venue.

These final analyses allow us to identify the key outlets leading the DAO conversation, the dominant topics they cover, and the main scholarly communities connected through these venues. Additionally, we examine which outlets and scholars belong to different components of these network and how they relate to the broader research community.

## 4. Results

### 4.1. Interdisciplinarity interactions between fields

The citation network analysis confirms that DAO-related knowledge creation heavily relies on boundary papers, which are articles cited by both Computer Science (CS) and Business and Economics (BE). Table 2 reports our results for each measure, leading to a Dominance Level ($D$) of 4.8, which indicates that interactions between the two fields are primarily mediated through boundary papers. Additionally, our linkage factor suggests a mutual exchange of knowledge, though CS contributes more to BE than vice versa.

**Table 2: Interdisciplinary interactions measures**

| Measure | Value | Interpretation |
|---|---|---|

7 Labels from different queries sharing one or more keywords, or labels depicting the same processes at different levels (e.g., blockchain/DAO) were considered consistent. Consistency tables can be found in Appendix D.

| | | |
|---|---|---|
| Strength Through Boundary Papers | 0.26 | Ratio of boundary papers (shared by both fields) to the total number of unique papers in the two networks. |
| Strength Through Cross-Disciplinary Arcs | 0.06 | Ratio of direct cross-field citations to the total number of edges in the merged network. |
| Dominant Mode of Interaction | 4.8 | Ratio of boundary papers to direct cross-disciplinary citations, indicating boundary papers as the predominant mode of interdisciplinary interaction. |
| Overall strength of interaction | 0.32 | Combined interaction strengths from both boundary papers and cross-disciplinary citations. |
| Mutual contribution between fields | BE → CS: 0.75<br>CS → BE: 1.33 | Ratio of net knowledge flow between the two fields. A higher value from CS to BE suggests greater knowledge transfer from Computer Science. |

A *D* value below 1 would suggest that fields interact primarily through direct cross-citation rather than boundary papers. While cross-disciplinary citations do exist, boundary papers dominate, indicating a shared knowledge space that both fields recognize as relevant.

Moreover, interactions between CS and BE are reciprocal, confirming that DAO research exhibits interdisciplinary qualities rather than merely multidisciplinary ones. Multidisciplinarity would imply a more unidirectional knowledge borrowing, where one field (typically BE) relies on concepts from another (CS) without meaningful integration. However, the data suggests that while business scholars are increasingly engaging with blockchain and DAO-related concepts, CS scholars cite BE research as well – even if far less frequently.

If, as hypothesized, interdisciplinary exchanges are still largely case-based and instrumental or methodological interactions, CS scholars would have little incentive to cite BE studies. Organizational research can integrate CS concepts – such as smart contracts or decentralization – but CS scholars have less reason to reference organizational theory. This pattern reinforces the early-stage nature of interdisciplinary research on DAOs, where organizational insights are not yet influencing CS knowledge production in significant ways.

Finally, to further explore the role of boundary papers, we identified the most cited papers (Table 3). Given our data collection strategy, which excluded works without a DOI, we deemed necessary to differentiate between considered references (those analyzed in our study) and total references (all works cited by the original papers). Also, citations correspond to the number of citations made from other studies involved in the network, and not to the overall citations made to the paper.

**Table 3: Most cited Boundary Papers**

| Title | Authors | Date | Citations | Considered references | Total references |
|---|---|---|---|---|---|
| Governance in the Blockchain Economy: A Framework and Research Agenda | Roman Beck, Christoph Müller-Bloch, John Leslie King | 2018 | 69 | 0 | 49 |
| Experiments in algorithmic governance | Quinn DuPont | 2017 | 66 | 0 | 65 |
| Decentralized Autonomous Organizations: Concept, Model, and Applications | Shuai Wang, Wenwen Ding, Juanjuan Li, Yong Yuan, Liwei Ouyang, Fei-Yue Wang | 2019 | 59 | 20 | 51 |
| Decentralized Autonomous Organization | Samer Hassan, Primavera De Filippi | 2021 | 55 | 17 | 60 |
| Bitcoin and the rise of decentralized autonomous organizations | Ying-Ying Hsieh, Jean-Philippe Vergne, Philip Anderson, Karim Lakhani & Markus Reitzig | 2018 | 38 | 7 | 35 |
| The invisible politics of Bitcoin: governance crisis of a decentralised infrastructure | Primavera De Filippi, Benjamin Loveluck | 2016 | 38 | 10 | 82 |
| Blockchains and the economic institutions of capitalism | Sinclair Davidson, Primavera De Filippi, Jason Potts | 2018 | 35 | 0 | 63 |
| Blockchain Governance—A New Way of Organizing Collaborations? | Fabrice Lumineau, Wenqian Wang, Oliver Schilke | 2020 | 35 | 36 | 180 |
| Blockchain challenges and opportunities: a survey | Zibin Zheng, Shaoan Xie, Hong-Ning Dai, Xiangping Chen and Huaimin Wang | 2018 | 31 | 0 | 86 |
| Defining Blockchain Governance: A Framework for Analysis and Comparison | Rowan van Pelt, Slinger Jansen,Djuri Baars, Sietse Overbeek | 2020 | 30 | 19 | 88 |

These results confirm the interdisciplinary nature of foundational DAO research, but also the initial stage in that exchanges remain at the *instrumental* and *methodological* level. Many highly cited works focus on case studies, general definitions, or applications, rather than theoretical integration. Thus, while boundary papers provide an initial view into interdisciplinary literature, they do not yet reveal deeper knowledge flows between disciplines.

### 4.2. Patterns of interdisciplinarity interaction

Our analysis identified two dominant interactions patterns: Boundary Triggered Emergence (33%) and Interdisciplinary Emergence (39%), which involve most boundary papers (see Table 4).

**Table 4: Boundary papers' modes of interaction**

| Mode of Interaction | % of Interactions | Interaction Structure |
|---|---|---|
| Interdisciplinary diffusion | 5% (CS to BE) | A → boundary paper → B |
| | 4% (BE to CS) | B → boundary paper → A |
| Circular diffusion | 6% (CS) | A → boundary paper → A |
| | 4% (BE) | B → boundary paper → B |
| Cross-fertilization emergence | 9% | A ← boundary paper → B |
| Boundary triggered emergence | 33% | A → boundary paper ← B |
| Interdisciplinary emergence | 39% | boundary paper → boundary paper |

Note: A paper can be involved in multiple interactions: the percentages of interactions for each mode are therefore an expression of the importance of that pattern on the total of interactions, and not on the total of papers involved in that specific pattern.

These results suggest that while scholars from established disciplines recognize DAOs as an emerging phenomenon, the field is primarily being advanced by a new interdisciplinary community. While the former group is driving Boundary Triggered Emergence, with scholars from both sub-fields referencing boundary papers, the latter is participating in Interdisciplinary Emergence, working at the intersection of sub-fields and citing other papers in that same intersection. Most boundary papers are involved in Interdisciplinary and Boundary Triggered Emergence (see Figure 1), meaning that the intersection of CS and BE serves as both a catalyst and a diffusion hub for new ideas.

Additionally, interdisciplinary research on DAOs exhibits bridging and restructuring properties: While citations among boundary papers are building new connections between disciplines and exploring the opportunities offered by the merging of different ideas and approaches, the references going from sub-fields to this liminary research are a sign of emerging opportunities for rethinking existing frameworks, theories and methodologies. Thus, while a flourishing interdisciplinary conversation is taking place, existing theories and methodologies are only beginning to adapt to DAO-related research

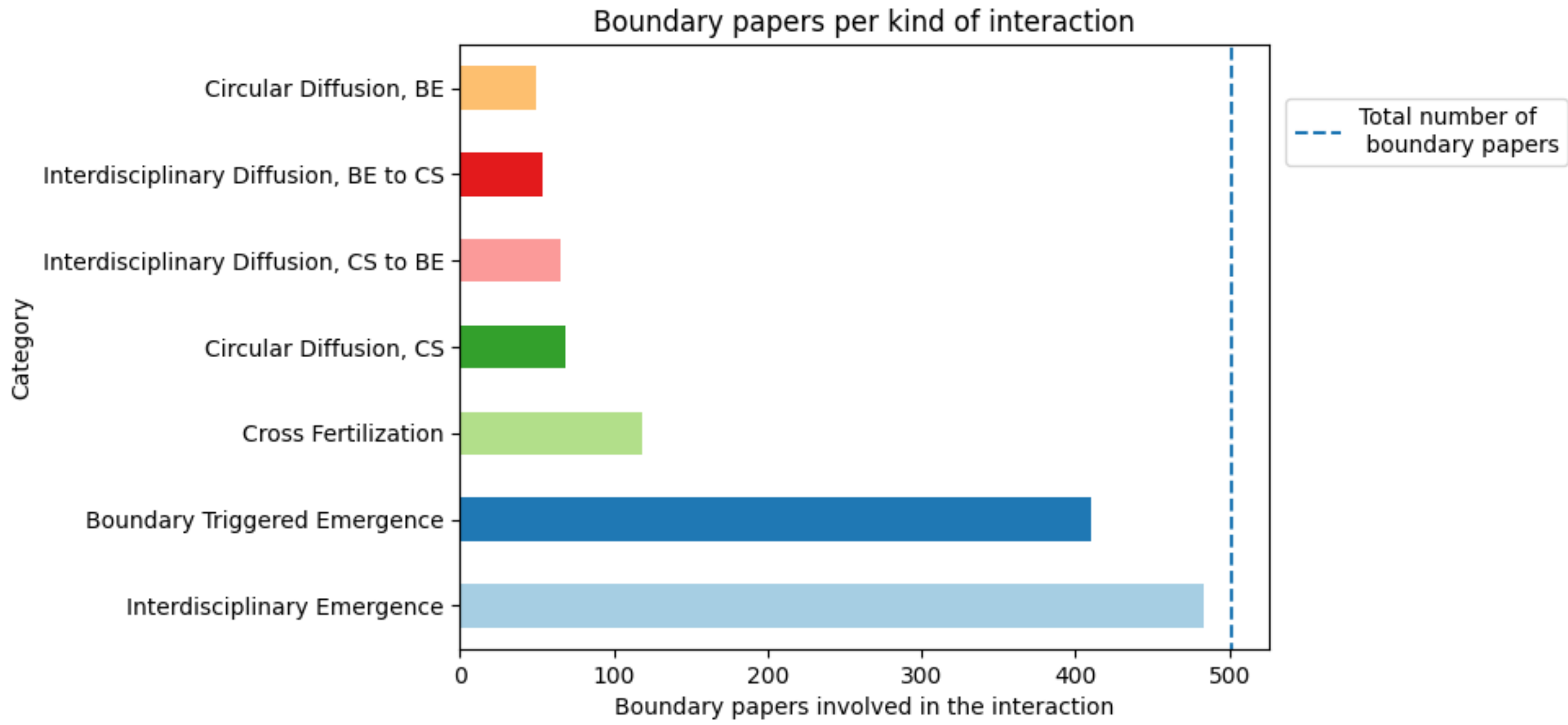

**Figure 1: Boundary papers involved in each kind of interaction**

### 4.3. Interdisciplinary paper communities

After testing the interdisciplinarity through citation networks, we further examined paper communities using network analysis and topic modeling. We used all edges belonging to the same mode of interaction to build a network for each possible interaction, identified communities based on Louvain modularity scores, extracted the main topic characterizing each of the networks, and normalized similar forms of the same topic to ensure comparability between networks.

The most connected network was Boundary Triggered Emergence (799 nodes, 2966 edges), followed by Circular Diffusion (957 nodes, 1826 edges), Interdisciplinary Diffusion (936 nodes, 1730 edges). Finally, we found Interdisciplinary Emergence (483 nodes, 1193 edges) and Cross-Fertilization Emergence (807 nodes, 1129 edges). Furthermore, only the Boundary Triggered Emergence, and the Interdisciplinary Emergence networks are formed by a single component, implying that all involved papers are somehow connected to one another, while components range from 3 to 5 for the other networks.

We will focus on the comparison of topics that appear in Interdisciplinary Emergence and Boundary Triggered Emergence (Figures 2 and 3), we observed that DAO-related discussions are concentrated in boundary papers, with DAOs being mentioned in three out of fourteen topics (including the one involving most papers) in the Interdisciplinary Emergence network, and that in BE and CS, DAOs appear in fewer topics (mostly governance & security). Nonetheless, looking at topics in the Boundary Triggered Emergence network, we can still observe DAOs' relevance for BE and CS, but with fewer topics (governance and security) and on a smaller degree.

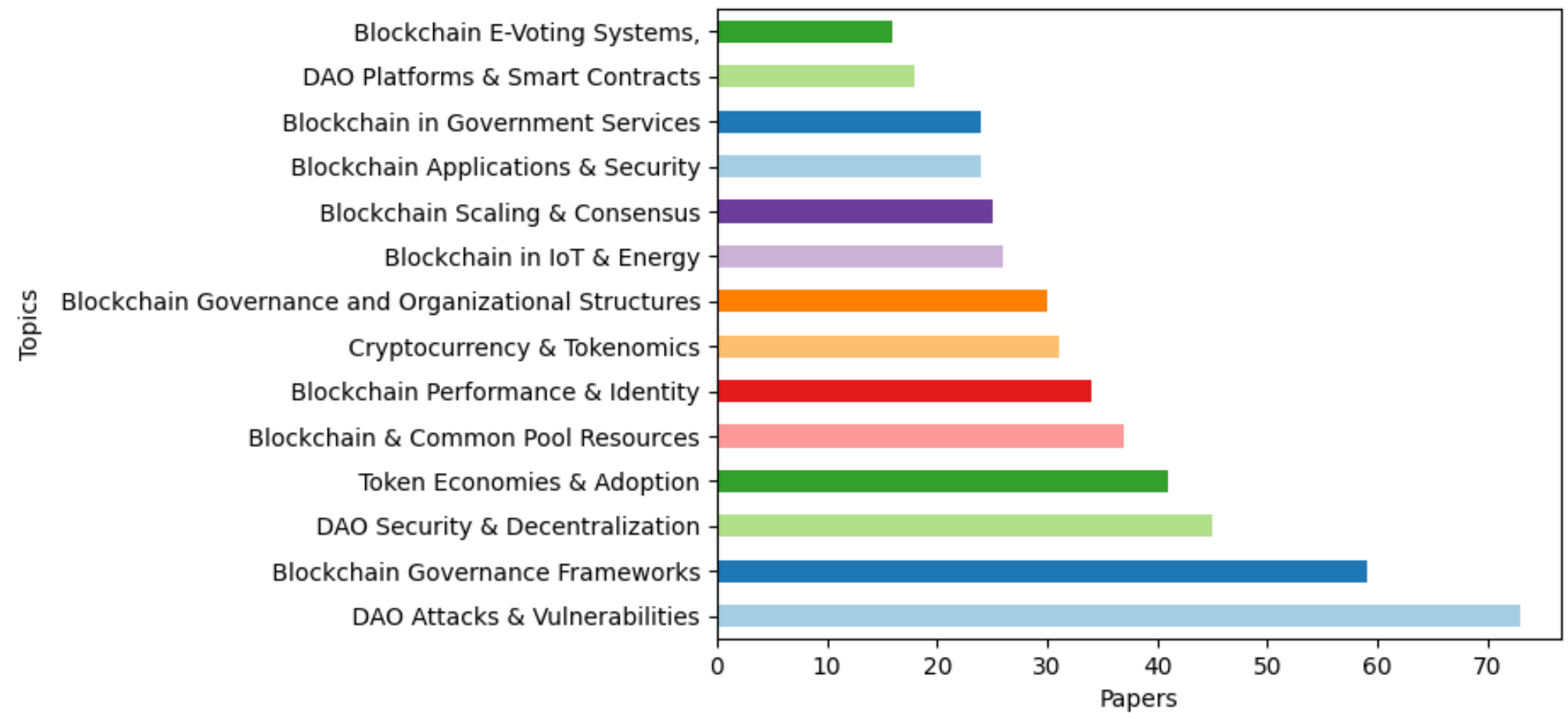

**Figure 2: Topics in the "interdisciplinary emergence" network**

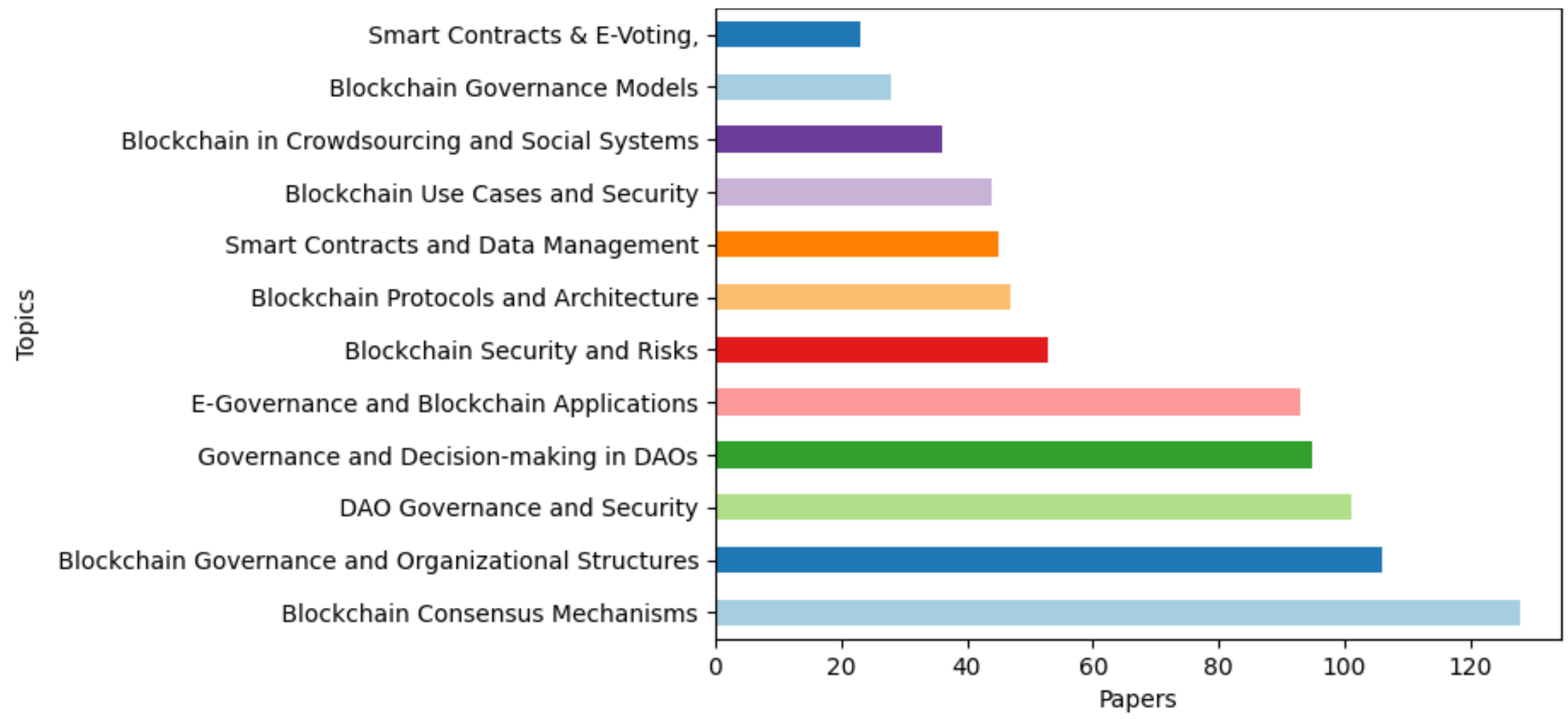

**Figure 3: Topics in the "boundary triggered emergence" network**

It should be noted that these papers are all somehow connected to one another, being part of the same (and only) component in these networks, suggesting a wide variety of topics are interacting in the current conversation. And while some topics fall in the CS domain (e.g., "Blockchain in IoT and Energy”), others are more related to BE (e.g., “Governance and decision making in DAOs”), or fall at the intersection of the two (e.g., “DAO governance and security”). Finally, while both networks exhibit a high variety of topics, one main difference is that while most topics in Interdisciplinary Emergence are built around domain-specific or practical applications (e.g., Blockchain applications and security, DAO attacks and vulnerabilities, Blockchain in government services), whereas topics in the Boundary Triggered Emergence network seem to be more high level and theoretically-oriented (e.g., Blockchain consensus mechanisms, Blockchain Governance and organizational structures, Blockchain governance models).

The nature of topics in Interdisciplinary Emergence aligns with the tendency of novel interdisciplinary communities to have practical applications as their primary focus (Frodeman, 2015), most likely leaning towards “methodological” and/or “instrumental” forms of interdisciplinarity rather than “theoretical” and/or “critical” ones (Klein, 2017). This is also supported by the presence of very technically oriented communities of papers (e.g., DAO attacks and vulnerabilities, Blockchain performance and identity).

Overall, these results suggest that, even if DAO research is vibrant and is already impacting existing domain knowledge in CS and BE, it is mostly relying on observational approaches, borrowing theories and methods from well-established knowledge domains. In order to further develop and grow, the field still needs to leverage existing exchanges to develop its own knowledge base, frameworks and theories. The following paragraphs will therefore be employed to identify relevant communities of practice and knowledge exchange hubs able to facilitate such processes.

### 4.4. Interdisciplinarity in publishing outlets and authors communities

We started our outlet analysis by inquiring where our boundary papers were published (Figure 4). Starting from the publications in the citation networks, we queried Scopus API to associate each paper with a journal or a conference and later aggregated the ones belonging to either different instances of the same conference or different conferences belonging to the same group (e.g., IEEE) and appearing only once or twice in the dataset.

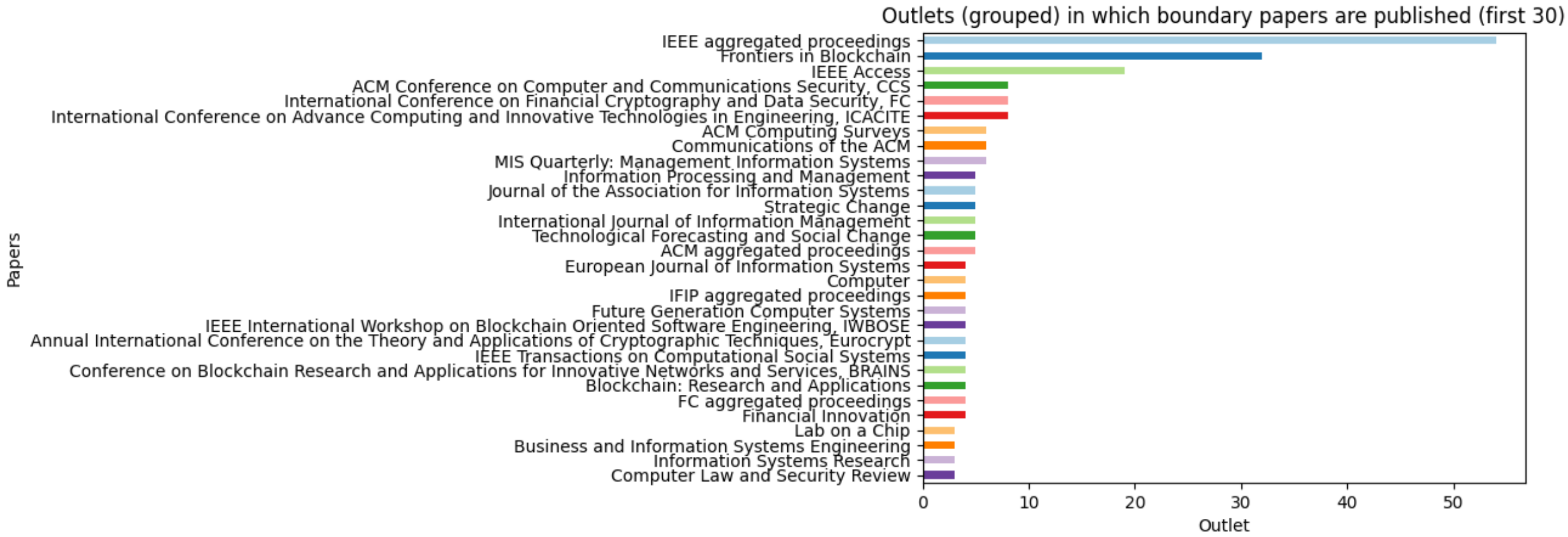

**Figure 4: Outlets hosting boundary papers**

Our results indicate a diverse publishing landscape, with IEEE conferences aggregating the majority of DAO-related papers, followed by *Frontiers in Blockchain* and *IEEE Access*. This distribution reflects a broad range of research communities engaging with DAOs, spanning computer science, business, and interdisciplinary outlets.

The *IEEE aggregated proceedings* confirm this variety, as they compile papers from numerous IEEE conferences, workshops, and symposiums that appeared only once or twice in our dataset. This suggests that at least 25-30 different IEEE conferences include DAOs or related topics among their areas of interest. Conversely, *Frontiers in Blockchain* has emerged as a specialized authority, attracting DAO-related research as a dedicated standalone journal.

To examine thematic coverage across these venues, we revisited our topic modeling results. As shown in Figure 5, the most dominant theme was "Blockchain governance and organizational structures", further confirming DAOs' central role among boundary papers. Other prominent topics included "DAOs attacks and vulnerabilities," "Blockchain consensus mechanisms," and broader discussions on frameworks and policies. These findings reaffirm DAOs' position within the larger blockchain discourse.

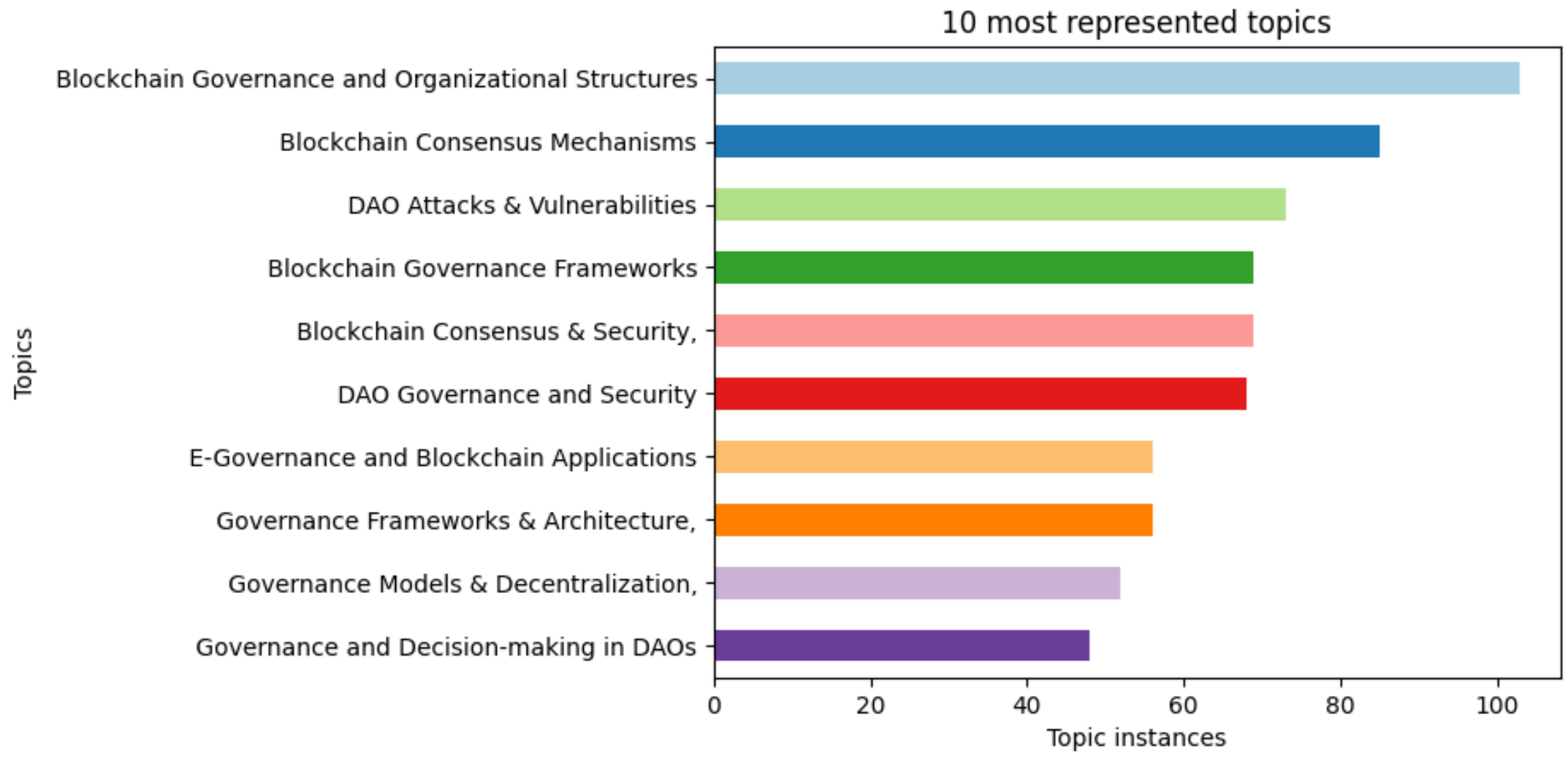

**Figure 5: Most represented topics**

It is essential to consider that, while our initial dataset specifically targeted papers discussing DAOs, the citation network was not artificially constrained in this way. Given that each paper in our dataset cited an average of 11.2 other papers in the network, and considering the limited existing literature on DAOs as a standalone topic, most of the network would be constituted by papers connected to the main topic but not directly addressing DAOs as the main focus of their analysis. The fact that DAOs are still the most represented topic in the communities of papers emerging from interactions with boundary papers is an additional confirmation of the interdisciplinary nature and importance of the phenomenon.

We next analyzed outlet and author networks (Figures 6 and 7, respectively).

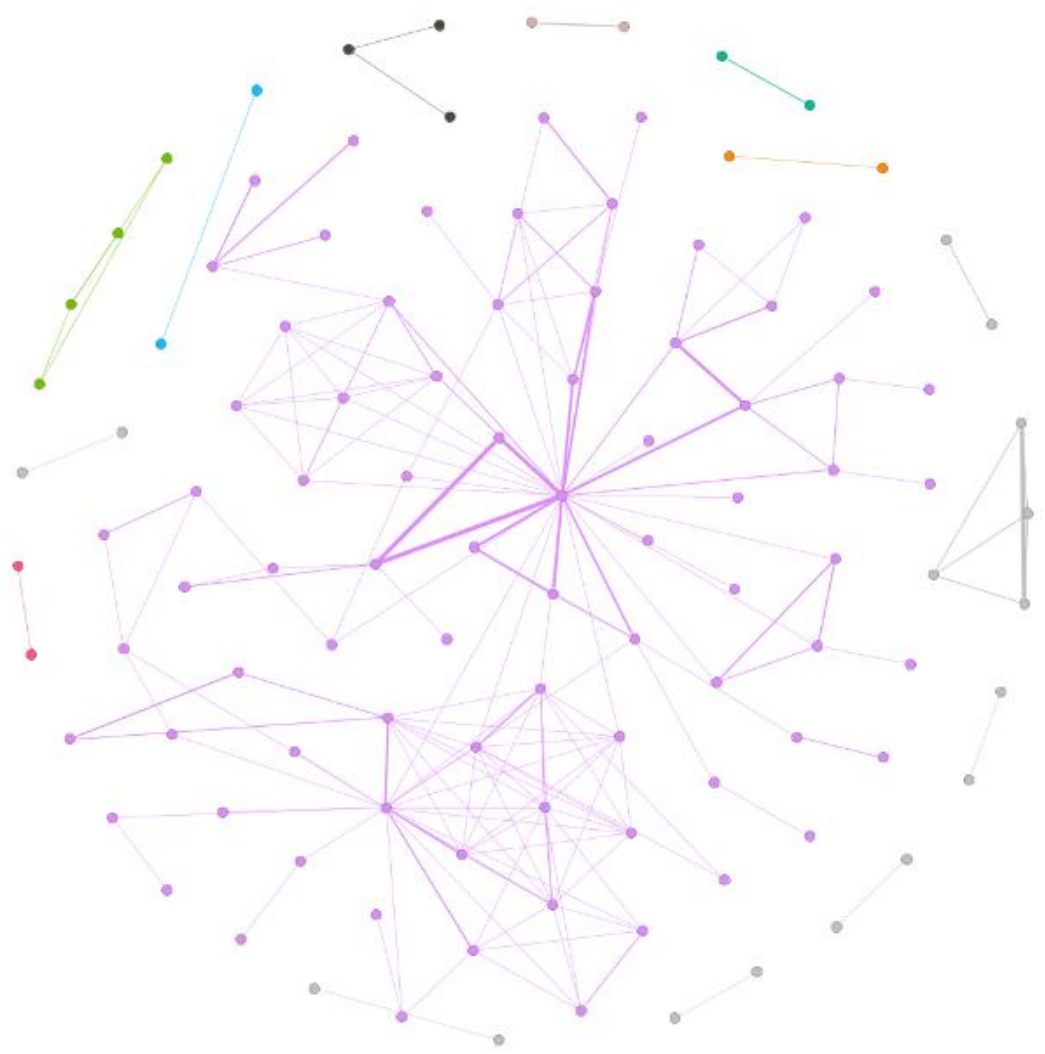

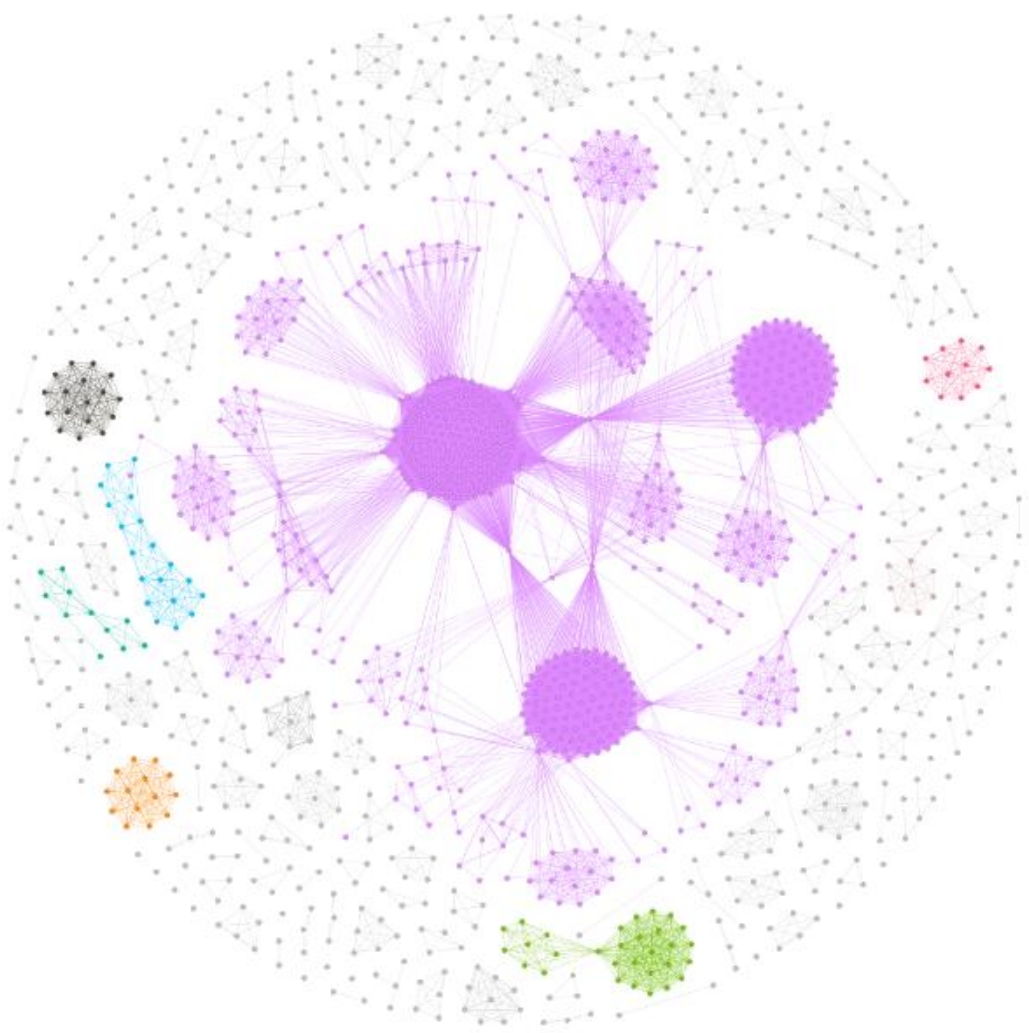

**Figure 6: Outlets network (colors identifying components) and Figure 7: Authors network (colors identifying components)**

Each resulting network showed multiple components, therefore indicating that, when looking at the structure of interactions through conferences and journals, we observe a very fragmented community. Nonetheless, a clear dominant component emerges from both representations and deserves to be further investigated.

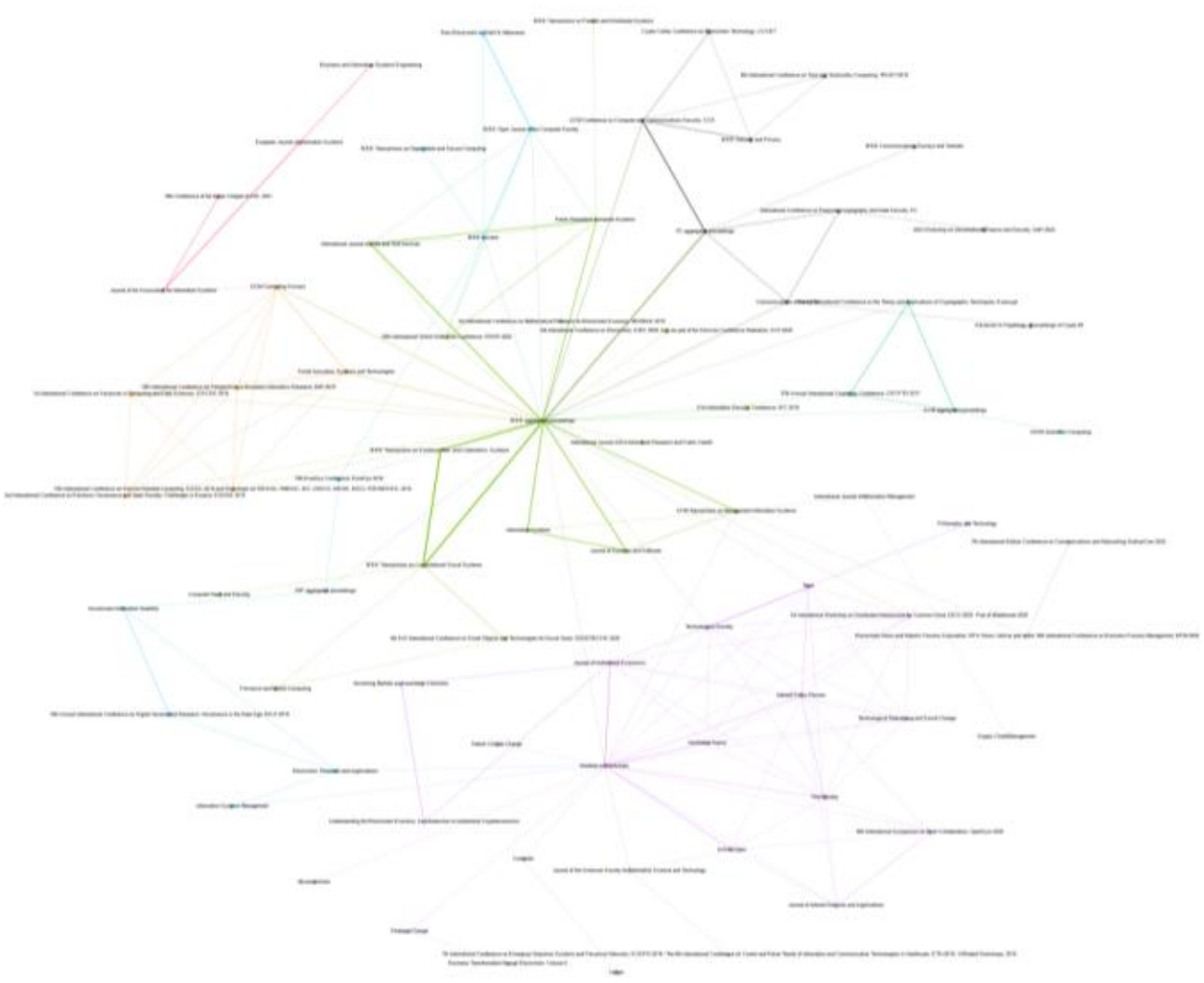

**Figure 8: Outlets network, main component (colors identifying modularity classes)**

Our outlet network analysis (Figure 8) revealed a fragmented research community but also identified a dominant subnetwork that merits further investigation. Specifically, the largest community consists of 78 outlets and 165 author connections (authors publishing in both outlets connected), comprising 70% of the entire network. While smaller clusters exist (ranging from 2 to 4 articles per cluster), this main subnetwork is the only widely connected component. The network includes 25 conference venues and 53 journal outlets. Among the conferences, four nodes correspond to aggregated proceedings from *IEEE*, *ACM*, *IPF*, and *FC*, while others include a mix of blockchain-specific and broader CS/business conferences (e.g., International Conference on Financial Cryptography and Data

Security, International Conference on Advances in Computing and Data Science, and International Conference on Perspectives in Business Informatics Research).

Given the size and structure of this main subnetwork, we assessed weighted degree, betweenness, and closeness centrality. Weighted degree centrality is the number of edges that each node has, and their weights; betweenness centrality measures the extent to which a node lies on paths between other vertices; closeness centrality is calculated as the reciprocal of the sum of the length of the shortest paths between the node and all other nodes in the graph. Nodes with high weighted degrees represent conferences with the most attendants in our dataset, while high betweenness can be interpreted as influence within the network, exerted by virtue control over information passing between others, and higher closeness centrality scores identify nodes closer to all other nodes. Given the size of the network, for each measure we only considered the five nodes with highest scores, reported in Table 5.

**Table 5: Centrality measures for main component of outlet network**

| Measure | Node 1 | Node 2 | Node 3 | Node 4 | Node 5 |
|---|---|---|---|---|---|
| Betweenness centrality | *IEEE aggregated proceedings* | *Frontiers in blockchain* | *ACM Transactions on Management Information Systems* | *ACM computing Surveys* | *Journal of the Association for Information Systems* |
| Closeness centrality | *IEEE aggregated proceedings* | *Frontiers in blockchain* | *Technology in Society* | *Topoi* | *ACM Transactions on Management Information Systems* |
| Weighted degree | *IEEE aggregated proceedings* | *Frontiers in blockchain* | *IEEE Transactions on Computational Social Systems* | *FC aggregated proceedings* | *IEEE Transactions on Systems, Man, and Cybernetics: Systems* |

*IEEE aggregated proceedings* dominate all measures, due to the broad participation across multiple IEEE conferences. *Frontiers in Blockchain* ranks second in all measures, making it the most influential standalone journal in DAO research. Other journals in the top three vary across centrality measures, indicating different forms of influence (e.g., bridging vs. proximity to other papers).

These findings suggest that *IEEE conferences* serve as key hubs where authors from various backgrounds converge, likely due to their broad thematic coverage. Interdisciplinary topics related to computational social systems and society are highly connected, suggesting these serve as natural intersections for interdisciplinary developments. *Frontiers in Blockchain* acts as a strong central node, providing a direct pathway for researchers to reach a diverse range of outlets and conferences.

Qualitatively, this allows us to provide some interesting observations: the first observation is about the bridging role performed by IEEE conferences, which seem to serve as an point of contact for authors then attending different conferences, probably thanks to the variety conferences offered by IEEE, which are also very thematically close to a multitude of other journals and conferences, and therefore attract the highest number of authors; topics related to computational social systems and society seem to be the also easily reachable from all other conferences, meaning they have the most shared set of attendees, probably because they serve as hubs for interdisciplinary developments; finally, the Frontiers journal seem to be a very central node, where most authors tend to publish, but that also enables them to connect to a wider set of outlets and conferences, given its high centrality in all measures.

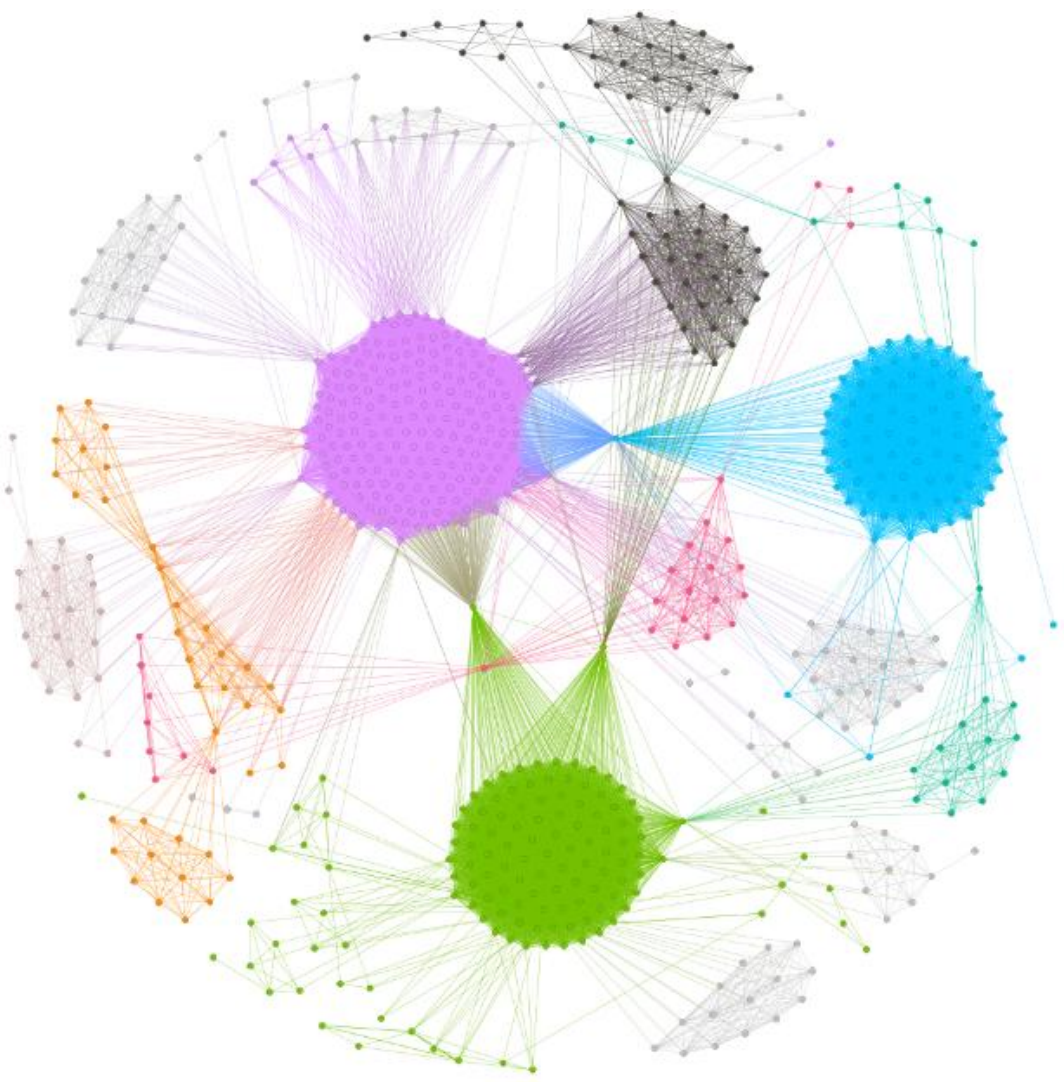

**Figure 9: Authors network, main component (colors identifying modularity classes)**

We next examined the author network (Figure 9) to identify key knowledge disseminators and inter-disciplinary connectors. The main point of interest in this network is that there are clearly nodes serving as bridges between highly interconnected communities (represented by different colors) that, should those nodes be removed, would be left disconnected from one another. Moreover, highly interconnected communities consist of scholars who publish in the same journals or attend the same conferences. In several cases, single authors – or a small group – serve as the only link between different subcommunities. This bridging effect highlights the importance of specific individuals in facilitating interdisciplinary exchanges. If these key authors were removed, certain subfields could become isolated, reducing knowledge flow across disciplines.

To further explore this, we mapped which conferences were connected by these bridging scholars (Figure 10).

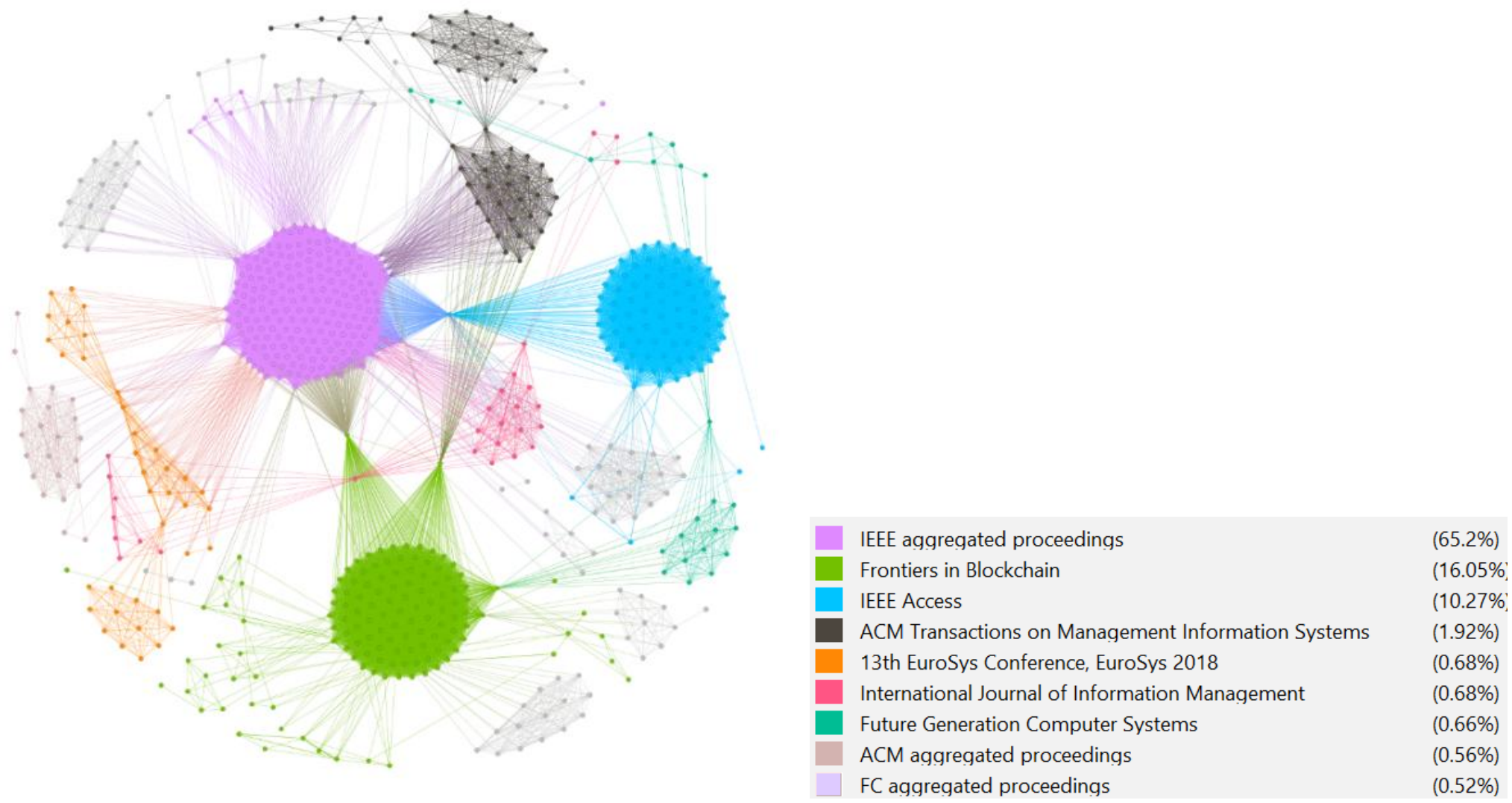

**Figure 10: Authors network, main component (colors identifying modularity classes/venues)**

By examining venue-based clustering in the author network, we identified a clear dominance of *IEEE Aggregated Proceedings*, which connects to almost every other conference via multiple authors. This confirms the *IEEE* cluster's role as a primary knowledge aggregator, likely due to its broad range of conferences covering interdisciplinary blockchain topics. *Frontiers in Blockchain*, while highly central, exhibits fewer direct connections to other clusters. Smaller communities appear well-connected among themselves and to *IEEE*, but connections between major venues depend on a small number of authors. The fragility of author-based connections underscores the need for stronger interdisciplinary collaborations, as existing exchanges rely on a limited number of key scholars. Put differently, large conferences like IEEE act as interdisciplinary hubs, bringing together researchers from different backgrounds, while stand-alone journals like Frontiers in Blockchain facilitate depth but may lack broad connectivity.

This aligns with the centrality measures calculated on the outlet network, but what is interesting here is that there are visibly few individuals maintaining those connections between biggest outlets, while smaller communities seem very well connected to each other and to the IEEE outlet. This informs us about the aggregating power of bigger conferences and at the same time identifies the real-life facilitators for interdisciplinary knowledge exchanges, highlighting the fragility of existing connections among different scholar domains.

## 5. Conclusions

This study highlights the inherently interdisciplinary nature of DAO research. We focus on the role of common knowledge (boundary papers) as crucial connectors between computer science and business economics, describing the forms and content of interdisciplinary interactions. By revealing how technological and organizational insights are interdependent, our work underscores the importance of integrated frameworks to examine the capabilities of DAOs in a more comprehensive way. Our findings demonstrate that, while significant progress has been made, the field remains fragmented, with a predominant emphasis on methodological and instrumental contributions rather than theoretical integration.

Our analysis examined interaction patterns, topic extraction, and collaboration networks, all of which indicate a highly interdisciplinary, yet still developing, field. The asymmetry in knowledge flows – with computer scientists contributing more to organizational research than vice versa – aligns with the type of interdisciplinarity in our findings: methods and tools are readily shared, while organizational theory has had less impact on computational research. The focus on case studies and applications, which emerges from our topic analysis, is characteristic of emerging interdisciplinary domains, thus confirming the field is still far from mature. By mapping key publication outlets and emerging themes, we provide a foundation for scholars to engage more effectively in interdisciplinary conversations. The dominant role of case studies and the centralization of research within a few key outlets are also typical of nascent research domains. Likewise, the limited presence of interdisciplinary theoretical contributions further reflects the field's early-stage development. Nevertheless, the prevalence of boundary-triggered emergence and interdisciplinary emergence patterns suggests that boundary papers are not only shaping discourse but also have the potential to influence and restructure pre-existing domains.

These findings have critical implications. For researchers, they underscore the need for frameworks that effectively bridge technical and organizational perspectives while providing a mapping of existing interdisciplinary hubs and conversations to leverage advancements made by interdisciplinary research. For practitioners, they highlight the transformative potential of DAOs in redefining governance models across industries, from decentralized finance to collaborative organizations. Moreover, our results indicate that DAOs and blockchain-based technologies could significantly reshape domain knowledge, potentially expanding and modifying existing theories on decentralization.

At the same time, our study has several limitations. Our reliance on bibliometric data may overlook qualitative nuances in interdisciplinary interactions, and our dataset may underrepresent emerging research in niche venues, as it is limited to publications indexed in Scopus and Web of Science. Additionally, while the exclusion of grey literature was necessary for consistency in data retrieval, this decision may have restricted our analysis – particularly given the high volume of non-academic publications in this domain. Furthermore, while our quantitative approach is comprehensive, our analysis only partially accounts for qualitative aspects of interdisciplinary collaborations and does so inconsistently across different steps. Addressing these gaps presents a valuable opportunity for future research.

Looking ahead, establishing collaborations, creating shared vocabularies and designing integrated frameworks will be crucial for advancing research on DAOs. Furthermore, scholars interested in the nature of the conversation around DAOs could explore the qualitative dynamics of interdisciplinary collaboration as well as include applications in emerging domains. Expanding the methodological approach to include qualitative studies – such as interviews with researchers and practitioners – could provide valuable insights into the real-world dynamics of interdisciplinary DAO research. Additionally, incorporating grey literature into the analysis could offer new perspectives on the discourse related to DAOs and capture insights beyond traditional academic publications. Further investigating and strengthening cross-disciplinary engagement may contribute to unlocking the full potential of DAOs.

Ultimately, this study aims at guiding scholars toward more theoretically grounded discussions on DAOs, offering a map of existing knowledge exchanges. By identifying relevant themes, outlets, and trends, we seek to foster interdisciplinary dialogue and facilitate the development of interdisciplinary perspectives and approaches for a deeper understanding of this phenomenon. DAOs have the potential to revolutionize governance models across industries, but their effective implementation requires insights from both technical and organizational standpoints. We hope this analysis serves as a foundational step toward integrating these perspectives in a meaningful way.

**Appendix A – Terminology**

*Web3*, also known as Web 3.0, envisions a new phase of the World Wide Web that integrates decentralization, blockchain technologies, and token-based economics. Blockchain serves as the driving force behind Web3, enabling user-owned data storage and management – including money, ownership, and identities – through self-sovereign identity, rather than relying on centralized entities for storage and management.

*Decentralized Autonomous Organizations* (*DAOs*) are virtual entities composed of members (or shareholders) who can vote, allocate funds, and modify the organization's code. Governed through on-chain smart contracts, DAOs replace traditional third parties to create a legal structure without centralized authorities. Their fair and trustless nature has made them increasingly prominent in the Web3 ecosystem.

*Blockchain* enables the existence of DAOs and is classified as a Distributed Ledger Technology (DLT) – a system of immutable "blocks" where transactions are recorded using cryptographic hashes. In a blockchain-based system, digital transactions are verified and stored without the need for central authorities, eliminating the need for trust between parties. Activities conducted directly on the blockchain are termed "on-chain", whereas activities occurring outside the blockchain are referred to as "off-chain". To establish trust, all participants must agree on the state of on-chain data when transactions are included and confirmed, regardless of whether decision-making occurs on-chain or through external platforms (e.g., Snapshot).

DAOs implement their flat and decentralized structure on blockchain through *smart contracts* and decentralized applications (*DApps*), which enforce rules, agreements, and governance mechanisms. Participants can create, deploy, and execute smart contracts to perform a variety of functions. Once deployed, a smart contract automatically enforces pre-defined triggers and conditions, facilitating agreements between users and implementing complex business logic.

The execution of monetary policies and economic regulations within blockchain-based projects—using cryptocurrency incentives to promote specific behaviors – is known as "*token economy*" or *Tokenomics*.

The token economy encompasses key *consensus* mechanisms, such as value transfer and smart contract execution, alongside economic activities characterized by rapid information transmission, strict execution rules, and efficient implementation. Tokenomics empowers individuals by granting them digital control over their identities, credentials, work products, and assets.

Moreover, Tokenomics plays a crucial role in incentivizing participation within decentralized blockchain networks. Its *rewards* and *reputation-based* metrics encourage active engagement in DAOs. Governance tokens – or other incentive mechanisms – ensure that active DAO members, not just investors, influence decision-making, fostering a collaborative and community-driven environment.

## Appendix B – Keywords and queries

When searching Scopus and WebOfScience, we employed the following keywords:

1. Scopus API: Keywords: ["decentraliz* AND autonomous AND organiz*", "decentralis* AND autonomous AND organis*", "dao* AND *governance", "blockchain* AND *governance", "dao* AND web*", "p2p AND dao*", "peer-to-peer AND dao*", "p2p AND blockchain* AND organiz*", "peer-to-peer AND blockchain AND organiz*", "p2p AND blockchain* AND organis*", "peer-to-peer AND blockchain* AND organis*", "daos", "smart contract* AND governance AND blockchain"]
2. WoS API: ((((((((((((TS=(decentraliz* AND autonomous AND organiz*)) OR TS=(decentralis* AND autonomous AND organis*)) OR TS=(dao* AND *governance)) OR TS=(dao* AND *governance)) OR TS=(dao* AND web*)) OR TS=(p2p AND dao*)) OR TS=(peer-to-peer AND dao*)) OR TS=(p2p AND blockchain* AND organiz*)) OR TS=(peer-to-peer AND blockchain AND organiz*)) OR TS=(p2p AND blockchain* AND organis*)) OR TS=(peer-to-peer AND blockchain* AND organis*)) OR TS=(daos)) OR TS=(smart contract* AND governance AND blockchain)

We queried the Scopus API for the following fields:

- "TITLE-ABS-KEY("+key+") AND SUBJAREA(BUSI OR DECI OR ECON)"
- "TITLE-ABS-KEY("+key+") AND SUBJAREA(COMP)"

## Appendix C – Prompt

### Gpt O1:

""""

I am giving you a list of titles and abstracts from papers on Decentralized Autonomous Organizations (DAOs), divided into communities.

The list is structured with the title first, followed by '---', then the abstract, then '---', and finally the community ID.

The list is organized as follows: { file with descriptions }

I need you to provide me, for each of the communities, a fitting short description (3 to 5 words), that summarizes the knowledge contained in the abstracts and titles belonging to that community. In case the abstract was not provided, papers show a 'Not found' in the corresponding field. In that case, try to look for the paper online, and if you cannot find any further information, just ignore the abstract.

I then need you to associate the description to each communty. The kind of output I need is a list structured as follows:

Community number: 0 --- Description: *put the description here*,

Community number: 1 --- Description: *put the description here*

etc.

Please start numbering communities starting from 0.

""""

## Appendix D – Topics consistency

**Table 1 - Topics consistency in Boundary triggered emergence**

| Community N. | Reported topics | Consistency (whole network) | Consistency (reverse order) |
|---|---|---|---|
| 1 | Blockchain Consensus Mechanisms | Blockchain Protocols and Security | Blockchain Consensus and Decentralization |
| 2 | Blockchain Governance and Organizational Structures | Blockchain Governance Mechanisms | DAO Governance Frameworks |
| 3 | DAO Governance & Security | DAO Governance and Sustainability Issues | DAO Governance and Security |
| 4 | Governance and Decision-making in DAOs | Blockchain Governance | Blockchain Governance and Regulation |
| 5 | E-Governance and Blockchain Applications | Blockchain Policies and Regulations | Blockchain Crowdfunding and Regulation |
| 6 | Blockchain Security and Risks | Privacy and Smart Contract Security | Blockchain Applications and Security |
| 7 | Blockchain Protocols and Architecture | Consensus Mechanisms and Decentralization | Consensus Mechanisms and Token Distribution |
| 8 | Smart Contracts and Data Management | Decentralized Governance Dynamics | Blockchain Governance and Power Dynamics |
| 9 | Blockchain Use Cases and Security | Technical Blockchain Challenges | Public Sector Blockchain Governance |
| 10 | Blockchain in Crowdsourcing and Social Systems | AI and IoT Integration with Blockchain | Blockchain Governance and Management |
| 11 | Blockchain Governance Models | Blockchain Governance Frameworks | DAO Frameworks and Models |
| 12 | Smart Contracts & E-Voting | Blockchain Voting Systems | Blockchain |

**Table 2 - Topics consistency in Interdisciplinary emergence**

| Community N. | Reported topics | Consistency (whole network) | Consistency (reverse order) |
|---|---|---|---|
| 1 | DAO Attacks & Vulnerabilities | Governance and Regulation | Blockchain Challenges & Opportunities |
| 2 | Blockchain Governance Frameworks | Blockchain Governance | Blockchain Governance Framework |
| 3 | DAO Security & Decentralization | Blockchain Security and Attacks | Blockchain Security Protocols |
| 4 | Token Economies & Adoption | E-Voting and Authentication | Blockchain Implementation & Access Control |
| 5 | Blockchain & Common Pool Resources | Decentralized Commons and Governance | Blockchain for Commons & Sustainability |
| 6 | Blockchain Performance & Identity | Blockchain Consensus Mechanisms | Consensus Mechanisms & Scalability |
| 7 | Cryptocurrency & Tokenomics | Blockchain Applications and Governance Surveys | Cryptocurrencies & Tokenomics |
| 8 | Blockchain Governance and Organizational Structures | Organizational Governance and Trust | Blockchain Governance & Policy |
| 9 | Blockchain in IoT & Energy | Blockchain Applications and Integration | Health Data on Blockchain |
| 10 | Blockchain Scaling & Consensus | Token Economies and Finance | Blockchain in Health & IoT |
| 11 | Blockchain Applications & Security | Smart Contracts and Vulnerabilities | Smart Contracts Vulnerabilities |
| 12 | Blockchain in Government Services | DAO Governance and Blockchain Integration | Blockchain in Government |
| 13 | DAO Platforms & Smart Contracts | Blockchain Voting Systems | DAOs & Decentralized Applications |
| 14 | Blockchain E-Voting Systems | Blockchain Consensus and Security | Blockchain E-Voting Systems |